\documentclass[aps,pra, onecolumn,14 pts,a4paper]{revtex4}
 \pdfoutput=1

\usepackage{amssymb,amsmath,amsfonts,graphicx}
\usepackage{bm}
\usepackage{epsfig,color,times}
 \usepackage{bm,color}
\usepackage{graphicx}
\usepackage{amssymb}
\usepackage{epstopdf}

\begin{document}

\title{ Conservation of the circulation for the Euler and Euler-Leray equations}

\author{Jean Ginibre$^{1}$, Martine Le Berre$^2$ and Yves Pomeau$^{3}$}
\affiliation{$^{1}$Laboratoire de Physique Th\'{e}orique (CNRS UMR 8627), Universit\'{e} de Paris-Sud, 91405 Orsay Cedex, France
\\$^2$Intitut des sciences mol\'{e}culaires d'Orsay (CNRS UMR 8214), Universit\'{e} de Paris-Sud, 91405 Orsay Cedex, France
\\ $^3$LadHyX (CNRS UMR 7646), Ecole Polytechnique, 91128 Palaiseau, France.}

\date{\today }

\begin{abstract}
It is well-known that the circulation of the velocity field of a fluid along a closed material curve is conserved for any solution of the Euler equation. We offer a slightly more explicit proof of that fact than that generally found in the literature. We then rewrite that property in terms of the rescaled variables and functions leading to the Euler-Leray equations and appropriate for studying self-similar solutions. We finally discuss the implications of the conservation of circulation on the existence of such solutions.

\end{abstract}

\maketitle

\section{Introduction}

   In this paper we reconsider the question of the conservation of circulation (CC) for the solutions of the Euler equation, a result known as Kelvin Theorem. We first rewrite the usual proof \cite{ LL} of that result in a slightly more explicit form than that given in \cite{ LL}. It then becomes immediately apparent that (i) the proof holds independently of the space dimension, and that (ii) not only does the Euler equation imply CC, but the converse is also true, namely (for sufficiently regular velocity fields), CC is equivalent to the Euler equation.
 Incompressibility  is not assumed  and  plays no role in that  question.

   We next rewrite the previous results in terms of suitably time rescaled variables and functions (see (\ref{eq:defX})-(\ref{eq:defG}) below) appropriate for studying self-similar solutions. Under a suitable condition on the scaling parameters (see (\ref{eq:alph-bet})), the Euler equation becomes equivalent to the time dependent Euler-Leray (TEL) equation (\ref{eq:TEL}) \cite{Yves}  \cite{giga}. We rewrite the circulation of the velocity field in terms of the new variables (see (\ref{eq:Gamt}) (\ref{eq:Gamtilda})) and we verify by a direct computation in those variables that CC is also equivalent to the TEL equation. 
   
   We finally turn to the case of self-similar solutions, associated with solutions of the stationary Euler-Leray (SEL) equation (\ref{eq:SEL}) and we investigate whether CC implies additional constraints on those solutions. For orientation, we first consider the well understood case of $L^2$-norm or equivalently (kinetic) energy conservation for the incompressible Euler equation. In that case, one sees easily that no $L^2$valued self-similar solution can exist unless the dilation parameters satisfy a suitable  $L^2$compatibility condition. 
   
   The case of CC for the Euler equation is more delicate. We can easily identify a CC compatibility condition for the dilation parameters (namely $\alpha=\beta$ in  (\ref{eq:defX}) (\ref{eq:defU})) under which no additional constraint occurs for the self-similar solutions. However we leave it as an open question whether the expected complementary result holds, namely whether for non CC compatible dilation parameters (namely $\alpha \ne \beta$), the self-similar solutions must have vanishing circulation, or equivalently be gradients.

   \section{Conservation of circulation}
    \label{sec:Euler eqs} 
    
   In this section, we rewrite the usual proof \cite{ LL} of the conservation of circulation (CC) for solutions of the Euler equation. We assume that all the functions written below are sufficiently regular for the subsequent computations to make sense. Let  $n$ be the space dimension. Let $u$ be a vector field defined on a domain  $D\times I$ of $R^{n} \times R$ , namely
      \begin{equation}
u :   (x,t) \; \; ( \in D\times I) \to u(x,t) \;\;  \in {R}^{n} 
\textrm{.}
\label{eq:u}
\end{equation}
A time dependent closed curve $ {\mathcal{C}} $ is the image of a circle, given by a continuous function $ \gamma$ depending on an angle variable $\theta$ and on time, namely
\begin{equation}
\gamma :   (\theta,t)\;\;  ( \in  [0,2\pi]\times I) \to \gamma(\theta,t) \in D
\textrm{,}
\label{eq:gamma}
\end{equation}
and such that $\gamma(0,t)=\gamma(2\pi,t)$. One can extend $\gamma$ to a continuous  $2\pi$  periodic function of $\theta$ .
The vector tangent to the curve ${\mathcal{C}} $ is  $\partial \gamma(\theta,t)/ \partial \theta$ by definition.
The circulation  of the vector field  $u$ along  the curve ${\mathcal{C}} $ at time $t$ is defined as the integral
   \begin{equation}
\Gamma(t)= \int_{0}^{2\pi} \mathrm{ d }\theta \; u(\gamma, t) \cdot \partial_{\theta} \gamma(\theta,t)
\textrm{.}
\label{eq:circ}
\end{equation}
We now assume that $u$ is the velocity field which makes  $\gamma$ evolve. That is expressed by the following  relation between $u$ and  the time derivative of $\gamma$,
\begin{equation}
\partial_{t}\gamma(\theta , t)= u(\gamma,t)
\textrm{.}
\label{eq:pg}
\end{equation}
We compute the time  derivative of the circulation
\begin{equation}
\partial_{t}\Gamma( t)=  \int_{0}^{2\pi}  \mathrm{ d }\theta  \; \left(   \partial_{t} u_{i}(\gamma,t) \partial_{\theta}\gamma_{i}(\theta,t) +    \partial_{j} u_{i}(\gamma,t) \partial_{t}\gamma_{j}(\theta,t) \partial_{\theta}\gamma_{i}(\theta,t) +u_{j } (\gamma,t) \partial_{\theta}  \partial_{t} \gamma_{j}(\theta,t) \right )
\textrm{.}
\label{eq:dcirc}
\end{equation}
The last term in the RHS of (\ref{eq:dcirc})  can be rewritten as
$$u_{j } (\gamma,t) \partial_{\theta}  \partial_{t} \gamma_{j}(\theta,t)= u_{j }(\gamma,t)  \partial_{\theta} u_{j }(\gamma,t)= u_{j } (\gamma,t) \partial_{i} u_{j } (\gamma,t)  \partial_{\theta} \gamma_{i} = \frac{1}{2}\partial_{\theta} \vert u \vert ^{2}.$$
We obtain
\begin{equation}
\partial_{t}\Gamma( t)=  \int_{0}^{2\pi}  \mathrm{ d }\theta  \; \left(   \partial_{t}u_{i}(\gamma,t)  +   u_{j}(\gamma,t) \partial_{j}u_{i}(\gamma,t)  +  \frac{1}{2}\partial_{i} \vert u \vert ^{2}  \right )\partial_{\theta}\gamma_{i}(\gamma,t)
\textrm{.}
\label{eq:dcirc2}
\end{equation}
Now the circulation of a gradient along any closed curve vanishes, namely for any function $f(x,t)$, the following holds
\begin{equation}
\int_{0}^{2\pi} \mathrm{ d }\theta \;\;\partial_{i} f(\gamma, t) \partial_{i} \gamma(\theta,t)= \int _{0}^{2\pi} \mathrm{ d }\theta \; \; \partial_{\theta} f(\gamma, t) =0
\textrm{.}
\label{eq:fcontour}
\end{equation}
This implies that
\begin{equation}
\partial_{t}\Gamma(t)= \int_{0}^{2\pi} \mathrm{ d }\theta \; v(\gamma, t) \cdot \partial_{\theta} \gamma(\theta,t)
\textrm{,}
\label{eq:circv}
\end{equation}
where  
\begin{equation}    
v=\partial_{t} u + (u \cdot \nabla) u
\textrm{,}
\label{eq:defv}
\end{equation}
namely the time derivative of the circulation of $u$ is the circulation of $v$. The CC of $u$ along any closed curve is therefore equivalent to the vanishing of the circulation of $v$ along any such curve, which is in turn equivalent to the fact that $v$ is a gradient. That is equivalent to the fact that there exists a function 
   \begin{equation}
 w: (x,t)  \; \; ( \in D\times I) \to w(x,t) \;\;  \in R
\textrm{,}
\label{eq:w}
\end{equation}
such that 
  \begin{equation}
\partial_{t} u +(u \cdot \nabla)u +\nabla w=0
\textrm{.}
\label{eq:euler}
\end{equation}
This proves that conservation of circulation for a (sufficiently regular) velocity field is equivalent to the equation (\ref{eq:euler}) for that field. Note that the incompressibility condition  $\nabla u=0$  has not been assumed and plays no role in the derivation of that result.
  
   In the compressible case, namely without imposing the incompressibility condition, the Euler equation takes the standard form
   \begin{equation}
 \rho \left (\partial_{t} u + (u \cdot \nabla)u \right )+ \nabla p = 0                                           
\textrm{,}
\label{eq:euler-rho}
\end{equation}
where  $\rho$  is the (possibly space time dependent) density of the fluid and   $p$  is the pressure. There is no reason for conservation of circulation to hold for a velocity field satisfying (\ref{eq:euler-rho}) unless  $\rho^{-1} \nabla p$ is a gradient. Therefore we assume that $\rho^{-1} \nabla p$ is a gradient. That assumption is supported by a thermodynamical argument in \cite {LL} and is easily seen to be satisfied if $p$ is a function of $\rho$, as the consequence of an equation of state. Under that assumption, one can define $$ \nabla w = \rho^{-1} \nabla p,$$
and (\ref{eq:euler-rho}) becomes identical with (\ref{eq:euler}). In the special case of incompressibility, which we do not assume here, $\rho$  is constant and can be normalized to $ \rho = 1$, so that $w = p$ in (\ref{eq:euler}). For simplicity of the exposition, we shall  refer to (\ref{eq:euler}) as the Euler equation in all that follows. We do not assume incompressibility, except where explicitly stated in part of section IV below.

 \section{ Conservation of circulation in rescaled variables }
    \label{sec:circ-scale} 
    In this section, as a preparation to the study of self-similar solutions of the Euler equation, we rewrite the results of the previous section in terms of appropriate rescaled variables and functions \cite{Yves} \cite{giga}. Aiming at solutions which concentrate at time zero, we take $t$ negative and for real (preferably nonnegative) $\alpha$ and $\beta$ , we define $X$, $\tau$,  $U$ and $W$ by
 \begin{equation}
X =\vert t \vert^{-\beta} x \;  \; , \; \;   \tau =-\ln\vert t \vert
\label{eq:defX}
\end{equation} 
                                                                                                                                                                                                                                                                                                                                                                                                                                                                                                                                                                               \begin{equation}
 u(x,t)=\vert t \vert^{-\alpha} U ( X, \tau )
\label{eq:defU}
\end{equation} 
\begin{equation}
 w(x,t)= \vert t \vert^{- 2 \alpha} W(X,\tau)
 \textrm{.}
 \label{eq:defW}
\end{equation} 
To any closed curve paramerized by a function $\gamma(\theta,t)$ we associate a rescaled closed curve parametrized by the function $G(\theta, \tau)$ defined by
\begin{equation}
    \gamma(\theta,t) = \vert t \vert^{\beta} G(\theta, \tau)
\textrm{.}
\label{eq:defG}
\end{equation}
  In terms of the variables  $X$ and $\tau $, and under the condition 
\begin{equation}
\alpha+\beta=1
\textrm{,}
\label{eq:alph-bet}
\end{equation}
 which we assume from now on, the  various terms in (\ref{eq:euler}) are  respectively given by
$$\partial_{t} u=\vert t \vert^{-\alpha-1}(\partial_{\tau}U + \alpha U + \beta(X \cdot \nabla) U) (X,\tau)$$
  $$(u\cdot \nabla)u= \vert t \vert^{-\alpha-1}  \left ( (U \cdot \nabla ) U \right )(X,\tau)$$
  and 
$$\nabla w(x,t)= \vert t \vert^{-\alpha-1} \nabla W(X,\tau).$$ 
Therefore
   \begin{equation}
\partial_{t} u +(u \cdot \nabla)u +\nabla w = \vert t \vert^{-\alpha-1}\left ( \partial_{\tau} U +\alpha U + \beta (X \cdot \nabla)U+ (U \cdot \nabla ) U  +\nabla W \right )(X,\tau)
\textrm{,}
\label{eq:self-sol}
\end{equation}
so that the Euler equation (\ref{eq:euler}) is equivalent to the following time dependent Euler-Leray (TEL) equation for U 
    \begin{equation}
\partial_{\tau} U +\alpha U + \beta (X \cdot \nabla)U+ (U \cdot \nabla ) U  +\nabla W=0
\textrm{,}
\label{eq:TEL}
\end{equation}
which is a generalization of the usual stationary Euler-Leray (SEL) equation, 
 \begin{equation}
\alpha U + \beta (X \cdot \nabla)U+ (U \cdot \nabla ) U  +\nabla W=0
\textrm{.}
\label{eq:SEL}
\end{equation}
 If $U$ satisfies (\ref{eq:TEL}), then $u$ defined by (\ref{eq:defU}) satisfies  the Euler equation (\ref{eq:euler}) and conversely. The circulation (\ref{eq:circ}) is conserved as proved in the previous section, namely $\partial_{t}\Gamma(t)=0$, with
    \begin{equation}
\Gamma(t)=  \vert t \vert ^{\beta-\alpha} \tilde{\Gamma}(\tau)
\label{eq:Gamt}
\end{equation}
and
\begin{equation}
\tilde{\Gamma}(\tau) =  \int_{0}^{2\pi}  \mathrm{ d }\theta \;\; U(G(\theta , \tau),\tau))\cdot \partial_{\theta} G(\theta,\tau) 
\textrm{.}
\label{eq:Gamtilda}
\end{equation}
  Since $\Gamma(t)$  is constant, $\tilde{\Gamma}$  is proportional to $ \vert t \vert^{\alpha-\beta} $ if $U$ is a solution of the TEL equation (\ref{eq:TEL}).
  
 It can be proved by a direct computation  \cite{sergio}  that $\Gamma$, as defined by (\ref{eq:Gamt}) (\ref{eq:Gamtilda}), is time independent if $U$ is a solution of the TEL equation, and actually, by the same argument as in the previous section, that the conservation of (\ref{eq:Gamt}) for any closed curve is equivalent to the TEL equation. The time derivative of $\gamma$ with respect to $t$, which is the value of the velocity on the  curve  ${\mathcal{C}}$, can be rewritten as
    \begin{equation}
 u(\gamma,t)=  \partial _{t} \gamma(\theta,t)= \vert t \vert ^{\beta-1} (-\beta G(\theta,\tau)+ \partial_{\tau} G(\theta,\tau))
\textrm{,}
\label{eq:dg}
\end{equation}
so that 
     \begin{equation}
\partial_{\tau} G(\theta,\tau)=\beta G(\theta,\tau)+ \vert t \vert ^{1-\beta-\alpha} U(G,\tau)
\textrm{.}
\label{eq:dG}
\end{equation}
 Under the condition (\ref{eq:alph-bet}) this yields
     \begin{equation}
\partial_{\tau} G(\theta,\tau)=\beta G(\theta,\tau)+  U(G,\tau)
\textrm{.}
\label{eq:dG1}
\end{equation} 
   We differentiate $\Gamma(t)$ given by (\ref{eq:Gamt}) (\ref{eq:Gamtilda}) with respect to time and use the fact that $\partial_{t} \tau= \vert t \vert ^{-1}$ for $t<0$. We obtain
\begin{equation}
\partial_{t}\Gamma(t)=  \vert t \vert ^{\beta-\alpha-1} \left( (\alpha-\beta) \tilde{\Gamma}(\tau) + \partial_{\tau} \tilde{\Gamma}(\tau)\right)
\textrm{.}
\label{eq:dtGam}
\end{equation}
  The time derivative of $\tilde{\Gamma}$ with respect to $\tau$ is
$$\partial_{\tau} \tilde{\Gamma}=\int \mathrm{ d }\theta \; \left( (\partial_{\tau}U_{i}+\nabla_{j}U_{i}\partial_{\tau} G_{j}) \partial_{\theta}G_{i} + U_{i}\partial_{\theta}\partial_{\tau} G_{i} \right).$$
 Inserting  (\ref{eq:dG1}) into this expression, we obtain
    \begin{equation}
\partial_{\tau} \tilde{\Gamma}=\int \mathrm{ d }\theta \; \left( \partial_{\tau}U_{i}+ (U \cdot \nabla)U_{i}+\beta U_{i}+ \beta(X \cdot \nabla) U_{i}  \right)   \partial_{\theta} G_{i}
\textrm{.}
\label{eq:dtGam1}
\end{equation}
Inserting (\ref{eq:dtGam1}) into (\ref{eq:dtGam}) yields
  \begin{equation}
\partial_{t}\Gamma(t)=  \vert t \vert ^{\beta-\alpha-1} \int \mathrm{ d }\theta \; \left( \partial_{\tau}U+ \alpha U + \beta (X \cdot \nabla) U + (U \cdot \nabla)U \right) \cdot \partial_{\theta}G 
\textrm{.}
\label{eq:dtGam2}
\end{equation}
The vanishing of (\ref{eq:dtGam2}) for any closed curve is equivalent to the fact that the integrand in the RHS is a gradient, which is equivalent to the fact that $U$ satisfies the TEL equation (\ref{eq:TEL}).

   One can rewrite (\ref{eq:dtGam2}) in a slightly different form, valid for any function $U(X,\tau)$ \cite{Rossi} by introducing the vorticity tensor $\Omega_{ij} = \nabla_{i}U_{j}- \nabla_{j}U_{i}$. We first recast (\ref{eq:dtGam2}) into the form
   \begin{equation}
\partial_{t}\Gamma(t)= (\alpha-\beta)  \vert t \vert ^{-1} \Gamma+
 \vert t \vert ^{-2\alpha} \int \mathrm{ d }\theta \; \left( \partial_{\tau}U+ \beta (1+X \cdot \nabla) U + (U \cdot \nabla)U \right) \cdot \partial_{\theta}G
\textrm{.}
\label{eq:dtGam3}
\end{equation}
We transform the last two terms of the integrand in (\ref{eq:dtGam3}) as follows 
$$(U+(X \cdot \nabla) U )_{i}= U_{i}+X_{j}\nabla_{j}U_{i}= 
U_{i}+X_{j}\nabla_{i}U_{j}-X_{j} \Omega_{ij} = \nabla_{i}(X\cdot U)-X_{j}\Omega_{ij}  $$ and 
$$\left( (U \cdot \nabla) U \right)_{i}= 
U_{j}\nabla_{j}U_{i}=  
 U_{j}\nabla_{i} U_{j}-   U_{j}\Omega_{ij}= 
 \frac{1}{2} \nabla_{i}\vert U \vert ^{2}   -U_{j}\Omega_{ij} . $$
 Since the gradient terms do not contribute to the integral in (\ref{eq:dtGam3}), we finally obtain
    \begin{equation}
\partial_{t}\Gamma(t)= (\alpha-\beta)  \vert t \vert ^{-1} \Gamma+
 \vert t \vert ^{-2\alpha} \int \mathrm{ d }\theta \; \left( \partial_{\tau}U_{i}-(\beta G _{ j} + U_{j})\Omega_{ij} \right) \partial_{\theta}G_{i}
\textrm{.}
\label{eq:dtGam4}
\end{equation}

Note in particular that all the terms in that equation vanish if $U$ is a gradient, since in that case  $\Gamma$ and $\Omega$ are identically zero while  $\partial_{\tau}U$, being also a gradient, does not contribute to the integral.

 \section{ Self-similar solutions }
    \label{self-sim} 
    
   In this section, we consider self-similar solutions of the Euler equation, namely solutions of the form (\ref{eq:defU}), always with $\alpha+\beta =1$ and with $\partial_{\tau}U=0$ , so that $U$ is a solution of the SEL equation (\ref{eq:SEL}). We want to investigate whether the conservation of circulation implies additional restrictions on those solutions. For orientation, we consider first the simpler case of the $L^{2}$-norm for the incompressible Euler equation, namely under the additional restriction 
  \begin{equation}
  \nabla \cdot u =0
  \textrm{.}
\label{eq:compress}
\end{equation}
In that case, it is well known that the $L^{2}$-norm is conserved.
For $u(x,t)$   defined in   (\ref{eq:u}) with $D=R^{n}$, we denote the squared $L^{2}$-norm by
   \begin{equation}
N(u)=\vert \vert u \vert\vert_{2}^{2}=\int_{R^{n}}  \mathrm{ d }x \vert u\vert^{2}
\textrm{,}
\label{eq:L2}
\end{equation}
which is twice the kinetic energy for the Euler equation. The  $L^{2}$-norm is covariant under the dilation   $\mathcal{D}_{\lambda}$ of positive parameter $\lambda$  defined by
\begin{equation}
(\mathcal{D}_{\lambda}u)(x,t) = \lambda^{-\alpha} u ( \lambda^{-\beta}x, \lambda^{-1}t )
\textrm{,}
\label{eq:dilat}
\end{equation} 
namely there exists an exponent $\delta$ such that 
  \begin{equation}
   N( \mathcal{D}_{\lambda}u) = \lambda^{\delta} N(u)
\textrm{.}
\label{eq:DL}
\end{equation} 
In the case considered here we have    
\begin{equation}
\delta= n\beta -2\alpha 
\textrm{.}
\label{eq:delta}
\end{equation}
  Let now $u$ be a self-similar solution of the incompressible Euler equation. Then $u(x,t)= (\mathcal{D}_{\vert t \vert}U)(x)$ and  in particular $U(x) = u(x,-1)$. Together with (\ref{eq:DL}) and $L^2$-norm conservation, this implies that for all $t$
\begin{equation}
N(u(t))= N(u(-1)) = N(U) = \vert t \vert ^{\delta} N(U) 
\textrm{.}
\label{eq:N2}
\end{equation}
  If $\delta = 0$  or equivalently if  $\alpha/\beta = n/2$, then (\ref{eq:N2}) is trivially satisfied and does not imply any additional constraint on self similar solutions. However if  $\delta\ne 0$, then (\ref{eq:N2}) implies that $N(U) = 0$ and therefore that $U = 0$ and $u = 0$. Therefore the incompressible Euler equation does not have (sufficiently regular) $L^2$valued self-similar solutions if $\alpha$ and $\beta$ do not take the $L^2$compatible values, namely such that $\alpha/\beta = n/2$. Under the condition $\alpha + \beta = 1$, the $L^2$compatible values become $\alpha = n/(n+2)$ and $\beta = 2/(n+2)$. In space dimension $n=3$ the $L^2$compatibility relation yields the values
  \begin{equation}
\alpha= 3/5 \qquad \beta= 2/5
\textrm{,}
\label{eq:sedov}
\end{equation}
originally deduced by Sedov and Taylor to describe blast waves.  The formation of  singularities  with exponents (\ref{eq:sedov}) was recently suggested  by the statistical analysis of the high speed velocity  flow recorded in the wind tunnel of Modane \cite{nous-coullet}.

  In order to gain some intuition on the behavior of self-similar solutions of the incompressible Euler equation, we assume that there exists a self-similar solution thereof such that the associated $U$ tends to zero at infinity in space as $R^{-k}$, where $R = \vert X \vert$, for some positive $k$. For large $R$, one expects the linear terms in the SEL equation (\ref{eq:SEL}) to be dominant. That requires that
 \begin{equation}
(\alpha + \beta (X\cdot \nabla)) R^{-k} = (\alpha - \beta k) R^{-k} =0
\textrm{,}
\label{eq:k}
\end{equation}
namely $k = \alpha/\beta$. The situation is then the following.
For the $L^2$compatible values $\alpha/\beta = n/2$, $k$ takes the (excluded) limiting value $k = n/2$ for $L^2$-norm finiteness at large distances.
For $\alpha/\beta > n/2$, the decay for large $R$ is sufficient to ensure $L^2$-norm finiteness at large distances. The lack of global $L^2$-norm finiteness is therefore expected to arise from non-$L^2$ local singularities. Physically that implies locally infinite energy, which is hardly acceptable. 
For $\alpha/\beta < n/2$, the decay for large $R$ is sufficiently slow to imply infiniteness of the $L^2$-norm, thereby leaving the possibility open for the existence of locally regular (at least locally $L^2$) solutions. Physically such solutions might be relevant, with locally finite energy, and even local energy in a suitably bounded region tending to zero as $t$ tends to zero. In order to ensure finiteness of the global energy, one would however need to consider solutions that are self-similar only in a bounded domain, but fail to remain so at large distances.

  The previous discussion applies with hardly any change to the Nonlinear Schr\"{o}dinger (NLS) equation
\begin{equation}
i\partial_{t} u + (1/2) \Delta u \pm \vert u\vert^{p-1}u = 0
\textrm{,}
\label{eq:NLS}
\end{equation}
for which the $L^2$-norm is well known to be conserved (for sufficiently regular solutions). Here $p > 1$, the function $u$ is complex valued and the relevant exponents $\alpha$  and $\beta$ are defined by $$(p-1)\alpha = 2\beta = 1.$$ By the same argument as for the incompressible Euler equation, the NLS equation does not have (sufficiently regular) $L^2$valued self-similar solutions unless $\alpha$ and $\beta$ take the $L^2$compatible values $\alpha = n/4$ and $\beta = 1/2$, namely unless $p$ takes the so called $L^2$critical value $p = 1 + 4/n$.

  We now  return to the Euler equation and to the case of the conservation of circulation (CC). On the basis of (\ref{eq:Gamt}) (\ref{eq:Gamtilda}), one is tempted to expect that the same argument as for the $L^2$-norm in the incompressible case applies to the circulation, with $N(u)$ and $N(U)$ replaced respectively by $\Gamma(t)$ and $\tilde{\Gamma}(\tau)$, and with $\delta = \beta - \alpha$. Indeed for the CC compatible values $\alpha = \beta$, both $\Gamma(t)$ and $\tilde{\Gamma}(\tau)$, which are then equal, are conserved by the general proof of Section II or III, so that CC does not imply additional constraints  on self-similar solutions in that case. In the complementary case  $\alpha \ne \beta$, in analogy with the case of the $L^2$-norm, one is tempted to expect that $\Gamma(t)$ and $\tilde{\Gamma}(\tau)$ vanish for self-similar solutions. Unfortunately the argument breaks down at that point because even for time independent $U$, and in contrast with  $N(U)$ which is then obviously time independent, the circulation $\tilde{\Gamma}(\tau)$ still depends on time through the curve, which evolves according to (\ref{eq:dG1}).Therefore in that case nothing more can be obtained from CC beyond the fact that $\tilde{\Gamma}(\tau)$ evolves in time as $\vert t \vert^{\alpha - \beta}$ as a consequence of (\ref{eq:Gamt}) and of the fact that $\Gamma(t)$ is constant by the general proof of Section \ref{sec:Euler eqs}  or \ref{sec:circ-scale}. This does not imply that $u$ and $U$ have vanishing circulation, or equivalently are gradients. Of course gradients have identically vanishing circulation and satisfy CC in a trivial way (see especially (\ref{eq:dtGam4}) and the subsequent comments). However we leave it as an open question whether for $\alpha \ne \beta$ self-similar solutions must have vanishing circulation, or equivalently be gradients.

\section*{Acknowledgement}
We are  very grateful to Christophe Josserand, Sergio Rica and  Maurice Rossi for stimulating discussions and correspondence on the subject matter of this paper.

{}
\end{document}